\documentclass[aps,prl, 
twocolumn,a4paper,
floatfix,superscriptaddress,showkeys,
]{revtex4-1}
\usepackage[utf8x]{inputenc}
\usepackage[pdftex]{hyperref}
\pdfoutput=1
\usepackage{bm,amsmath,comment,mathrsfs,datetime,
IEEEtrantools,
latexsym,amssymb,graphicx,xcolor,float}

\settimeformat{ampmtime}

\newcommand	{\bew}			{\begin{widetext} \begin{equation}}
\newcommand	{\eew}			{\end{equation} \end{widetext}}
\newcommand	{\be}			{\begin{equation}}
\newcommand	{\ee}			{\end{equation}}
\newcommand	{\bea}			{\begin{IEEEeqnarray}}
\newcommand	{\eea}			{\end{IEEEeqnarray}}
\newcommand	{\bs}			{\begin{split}}
\newcommand	{\eyz}			{\end{split}}
\newcommand	{\sch}			{Schr\"{o}dinger }

\newcommand	{\dd}			{\mathrm{d}}
\newcommand	{\tp}			{t^\prime}
\newcommand	{\eqr}		[1]	{eq. \eqref{#1}}
\newcommand	{\kett}		[1]	{|#1 \rangle}
\newcommand	{\braa}		[1]	{\langle #1|}

\newcommand	{\parnth}	[1]	{\left( #1 \right)}
\newcommand	{\sqbr}		[1]	{\left[ #1 \right]}
\newcommand	{\modsq}	[1]	{|#1|^2}
\newcommand	{\oter}		[2]	{|#1 \rangle \langle #2|}

\newcommand	{\forr}		[4]	{(#1,#2 \mid #3,#4)}

\begin{document}

\title{Quantum First-Passage Time:\\ Exact Solutions for a Class of Tight-Binding
Hamiltonian Systems}
\author{V. Ranjith}
\affiliation{Centre for Quantum Information and Quantum Computation,
Department of Physics, Indian Institute of Science, Bengaluru, 560012, India.\\
email: ranquest[AT]gmail.com}
\author{N. Kumar*}
\affiliation{*Author to whom all the correspondences should be addressed. \\
Raman Research Institute, Bengaluru, 560080, India.\\
email: nkumar[AT]rri.res.in}

\begin{abstract}
{The \sch integral-equation approach  {for calculating the} classical first-passage time (C-fpt) probability density is extended to the case of quantum first-passage time (Q-fpt). Using this extension, we have calculated analytically the Q-fpt probability density for a class of few-site/state tight-binding Hamiltonian systems, $e.g.,$ a qubit, as well as for an infinite 1D lattice. The defining feature of such a quantum system is that the passage across the boundary between a subspace ($\omega$) and its complement ($\bar{\omega}$) is through a unique pair of \textit{door-way} sites such that the first departure from (arrival at) $\omega$ corresponds to the first arrival at (departure from) $\bar{\omega}$.  The Q-fpt probability density so derived remains positive over the time interval in which it also normalizes to unity. These conditions of positivity and normalization define the physical time domain for the Q-fpt problem.  This time domain is found to remain finite for the few-site/state Hamiltonian 
systems considered here, which is quite unlike the case for the diffusive C-fpt problems. The \textit{door-way} sites and the associated Q-fpt probability density derived here should be relevant to inter-biomolecular/nanostructural electron transport phenomena.}
\end{abstract}



\maketitle

For a particle restricted to lie initially at a point $\nu$ inside a finite subspace $\bm \omega$ bounded by the surface $\bm \sigma$, the first-passage time ({\bf fpt}) is defined as the time of its first crossing of the boundary $\sigma$ (concomitant with its first arrival at the complementary subspace $\bm{\bar{\omega}}$) \cite{1,2,3,4,5,6}. Thus, for a classical ($\bm C$) stochastic system evolving probabilistically, one can conveniently introduce the \textit{restricted} ($\bm r$) probability $\bm {P^C_r\forr{\omega}{t}{\nu}{0}}$ that the particle remains confined to the subspace $\omega$ for all times $t^\prime$ with $0\leq t^\prime \leq t$. The associated C-fpt probability density is then necessarily given by $\bm{ P^C_{fp}\forr{\omega}{t}{\nu}{0}~\equiv ~-\left( \frac{\dd}{\dd t}P^C_r \forr{\omega}{t}{\nu}{0}\right)}$ \cite{4}. The parameters $\omega$ and $\sigma$ are to be chosen as appropriate to the specific physical realization of the problem. (It is to be noted here that the crossing of a 
boundary may, in general, connote crossing any sharply defined physical condition, $e.g.,$ of the first departure from, or the first arrival at the subspace of interest). The fpt-probability distribution is of interest for a number of physical, chemical and biological rate processes involving, $e.g.,$ the classical probabilistic (Kramers) escape over a potential barrier \cite{2,7}, the quantum tunneling through an activation barrier and the associated delay \cite{7,8,9,10,11,12,13,14,15,16}, the single-electron quantum transfer in nanostructures \cite{17} and biomolecules \cite{18}, and, of course, for the classic case of nuclear $\beta$-decay. For a classical stochastic process, $e.g.,$ physical diffusion, the fpt problem is well posed, and the problem has been solved variously and exhaustively \cite{1,2,3,4,5,6}, $e.g.,$ by  use of a perfect absorber $-$ a boundary condition that can be implemented mathematically through the well known Kelvin method of images \cite{1,2}.

An elegant approach to the classical fpt-probability density for a 1D continuum is the one based on the integral equation relating the \textit{restricted} ($r$) and the \textit{unrestricted} ($u$) probabilities ($P_r^C$ and $P_u^C$), as proposed by \sch \cite{4} (well before the appearance of the \sch wave equation of quantum mechanics), written symbolically as
\be \label{e1}
\begin{split}
P_u^C\forr{&\omega}{t}{\nu}{0} ~{\bm =}~ P_r^C\forr{\omega}{t}{\nu}{0} \\
&-\int_0^t \left( \frac{\partial}{\partial t^\prime} P_r^C \forr{\omega}{t^\prime}{\nu}{0} \right) P_u^C \forr{\omega}{t}{\bar{\omega}}{t^\prime} \dd t^\prime.
\end{split}  \ee

The essential logic underlying the above integral equation is simple: The first part of r.h.s. is the \textit{restricted} probability of never venturing beyond the subspace $\omega$ up to time $t$. The second part is the complementary probability that takes into account the \textit{multiple excursions} across the boundary $\sigma$  taking the particle to $\bar \omega$ at some time $0<\tp<t$, and then bring it back to $\omega$ within the remaining time $(t-\tp)$. This is an exhaustive reckoning of all the possible alternatives subsumed in the $P^C_u$ on the l.h.s of \eqr{e1}.

\begin{figure}
\centering
\includegraphics[height=77pt]{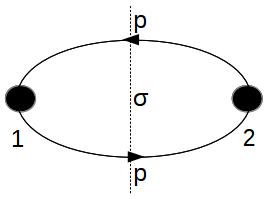} 
\caption{{\small Classical system of a particle hopping between the two sites 1$(\equiv\omega)$ and 2$(\equiv\bar{\omega})$; the dotted line depicting notionally the boundary $\sigma$, with $p$ the hopping probability rate.}}  \label{f1}
\end{figure}

Thus, $e.g.,$ for a classical 2-site hopping model with the subspaces (or rather the subsets) $\omega \equiv 1$ and $\bar{\omega} \equiv 2$ as depicted in Fig. \ref{f1}, the integral equation reads
\be \label{e2}
\begin{split}
P_u^C\forr{&1}{t}{1}{0}= P_r^C\forr{1}{t}{1}{0}\\
	   &-\int_0^t \dd t^\prime~ P_u^C \forr{1}{t}{2}{t^\prime}
	   ~\frac{\partial}{\partial t^\prime}P_r^C\forr{1}{t^\prime}{1}{0}.
\end{split}
\ee
The unrestricted probabilities involved here are nothing but solutions of the master equation for the system under appropriate boundary conditions. With this, it is straightforward to solve the integral equation, $e.g.,$ by a simple Laplace-transform \cite{6}. The time of first passage then refers to the time of first departure from the initial site 1, which is, of course, concomitant with the time of its first arrival at site 2. The integral equation (2) written for the case of C-fpt now sets the notation for the quantum case (Q-fpt) of a class of N-site systems where the boundary ($\sigma$) separating the subspace $\omega$ and its complement $\bar{\omega}$ is notionally a single point as depicted by the dotted vertical line in Fig. \ref{f1}.

Before we turn to Q-fpt, let us note that the Q-fpt problem has been generally viewed as ill-posed, and has, therefore, remained unsolved so far to the best of authors' knowledge (see however \cite{5}).  Reasons being: (a) inapplicability of Kelvin's method of images inasmuch as quantum mechanics does not admit perfect absorber \textit{sans} reflection \cite{6,19,20}; (b) essential difference between the Wiener (real) functional integral \cite{21} of classical stochastic mechanics and the Feynman (complex) path integral \cite{22} of quantum mechanics; (c) non-demolition continuous unobservability of quantum paths. (However, see \cite{23,24} for the possibility of continuous \textit{weak measurements}); and finally, (d) the question of time itself being an operator \cite{25,26,27,28}. (However, see \cite{29}).

In any case, the Q-fpt probability distribution is operationally well defined $-$ after all, for any physical event that may come to a pass, there must necessarily be a time of its first passage past the post ($i.e.,$ the marker or the boundary). To us, this seems to be a robust idea. The question then really is how it is to be calculated quantum mechanically. It is this central issue that we turn to now.
 
For the class of discrete (N-site) problems with only nearest-neighbor couplings as considered here, the subspace (or rather the subset) $\omega$ has a unique single site of departure to its complement $\bar{\omega}$, and $\bar{\omega}$ has a unique single site of arrival from  ${\omega}$.
(It is apt to call such a unique pair of sites bridging $\omega$ and $\bar{\omega}$ as the \textit{door-way sites} in general). Now, this uniqueness of connectivity eliminates any quantum interference of alternatives --other than those already contained in the \textit{unrestricted} probabilities $P^Q_u\forr{1}{t}{1}{0}$ and $P^Q_u\forr{1}{t}{2}{t^\prime}$. The latter would now replace their classical counterparts in \eqr{e2}, as the problem of quantum phase in connecting the \textit{restricted} and the \textit{unrestricted} probabilities at the point of return form $\bar{\omega}$ in \eqr{e2} now disappears. It is to be emphasized here that this crucial elimination of the interference of phase-alternatives is specific to the class of \textit{models with door-way sites} as considered here. Thus, the \sch integral equation for the classical case remains valid even for the quantum case, but with the proviso that the two \textit{unrestricted} probabilities referred to above are calculated quantum mechanically. 
The \textit{restricted} probability $P^Q_r$ is then determined \textit{implicitly} through the integral \eqr{e2} itself. We now turn to calculating the Q-fpt probability density in various cases.

First, we consider the two site/state quantum system (essentially a qubit) described by the
tight-binding Hamiltonian
\begin{equation} \label{e3}
\mathcal{H}_2 ~~=~~ \epsilon_1 \kett{1} \braa{1}  + \epsilon_2 \kett{2} \braa{2}  -\hbar\gamma
\kett{2} \braa{1}- \hbar \gamma^* \kett{1} \braa{2}  ,
\end{equation}
where $\kett{1}$ is taken to correspond to the subspace $\omega$, and $\kett{2}$ to the complement
subspace $\bar{\omega}$.  Here for simplicity of calculation, we will set the site energies
$\epsilon_1 = 0 = \epsilon_2$, and take $\gamma$, the tunneling matrix element, to be real with $\gamma = 1 = \hbar$. Here, the boundary ($\sigma$) is notionally a single point as depicted by a vertical dotted line in its classical analogue in Fig. \ref{f1}. For this quantum two-site system, the \sch integral equation reduces to
\begin{equation} \label{e4}
\begin{split}
P_u^Q\forr{1}{t}{1}{0} &= P_r^Q\forr{1}{t}{1}{0}\\
& -\int_0^t \dd t^\prime ~ P_u^Q\forr{1}{t}{2}{t^\prime} ~ \frac{\partial}{\partial t^\prime}
P_r^Q\forr{1}{t^\prime}{1}{0}.
\end{split}
\end{equation}

The structure of Eq.(4) embodies the following physical conditions:
\begin{itemize}
\item[(A).] The restricted probability \mbox{$P^Q_r\forr{1}{t}{1}{0}$} decreases monotonically from unity at $t=0$ to its first zero at some time $T$, which can even be infinite;
\item[(B).] It follows from (A) that the Q-fpt probability density
\mbox{$P^Q_{fp}\forr{1}{t}{0}{0}$}
~$\equiv$~
\mbox{$-\frac{\partial}{\partial t} P^Q_r\forr{1}{t}{1}{0}$}
is positive (but not necessarily monotonic), and is \textit{automatically normalized to unity} over the time interval
\mbox{$0 \leq t \leq T$}
, as
\mbox{$\int_0^{T} -\frac{\partial}{\partial t} P^Q_r \forr{1}{t}{1}{0}\dd t$}
~$=$~
\mbox{$P^Q_r\forr{1}{0}{1}{0}$}~
\mbox{$-~P^Q_r\forr{1}{T}{1}{0}$}~
$=~1~$; and,
\item[(C).] It follows from (A) and (B) that the time interval \mbox{$0 \leq t\leq T$} is the physical \textit{time domain of definition} for the Q-fpt problem.
\end{itemize}

The two \textit{unrestricted} probabilities in Eq. (4) are actually the results of the quantum unitary evolution as described by the time-dependent \sch equation. They are obtained as follows :
\be \label{e5}
P_u^Q \forr{1}{t}{1}{0} ~=~ \modsq{ \braa{1}  e^{-i \mathcal{H}_2t} \kett{1} } ~=~\cos^2 (t),
\ee
\be \label{e6}
P^Q_u \left(1, t\mid 2,t^\prime\right) ~=~ \modsq {\braa{1}e^{-i \mathcal{H}_2(t - t^\prime)}
\kett{2} } ~=~ \sin^2 (t - t^\prime)~.
\ee

This is, of course, not the case for the \textit{restricted} probability $P_r^Q(t)$. But, as
discussed earlier, the whole point of our treatment is that this \textit{restricted} probability
$P_r^Q(t)$ gets determined \textit{implicitly} through the \sch integral equation (4) in
terms of the \textit{unrestricted} probabilities $(P_u^Q)$ for the class of quantum systems with the door-way sites as considered here.

\begin{figure}
\centering
\includegraphics[width=.45\textwidth, clip=true,trim=0pt 0pt 0pt 25pt]{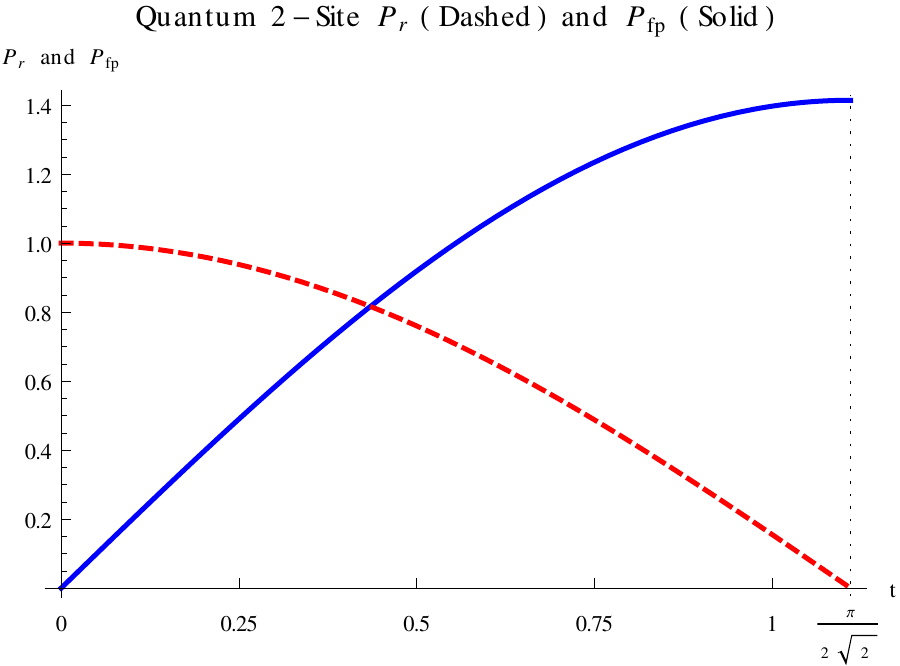}
\caption{\small{Quantum 2-site system: plots of the restricted probability
$P_r^Q\forr{1}{t}{1}{0}$ (dashed line) and the first passage probability density
$P^Q_{fp}\forr{1}{t}{1}{0}$ (solid line) against $t$, with $T = \frac{\pi}{2\sqrt{2}}$.}} \label{f2}
\end{figure}

Substituting from eqs.(5) and (6) into \eqr{e4} yields, after some algebra involving Laplace transforms, the required solutions as given below and plotted in Fig.\ref{f2} .
\be \label{e7}
P_r^Q\forr{1}{t}{1}{0}~~=~~\cos(\sqrt{2}~ t),
\ee
\be \label{e8}
P^Q_{fp}\forr{1}{t}{1}{0} ~~=~~ \sqrt{2}~~\sin(\sqrt{2}~t) .
\ee

\begin{figure}
\centering
\includegraphics[width=.45\textwidth, clip=true,trim=0pt 0pt 0pt 14pt]{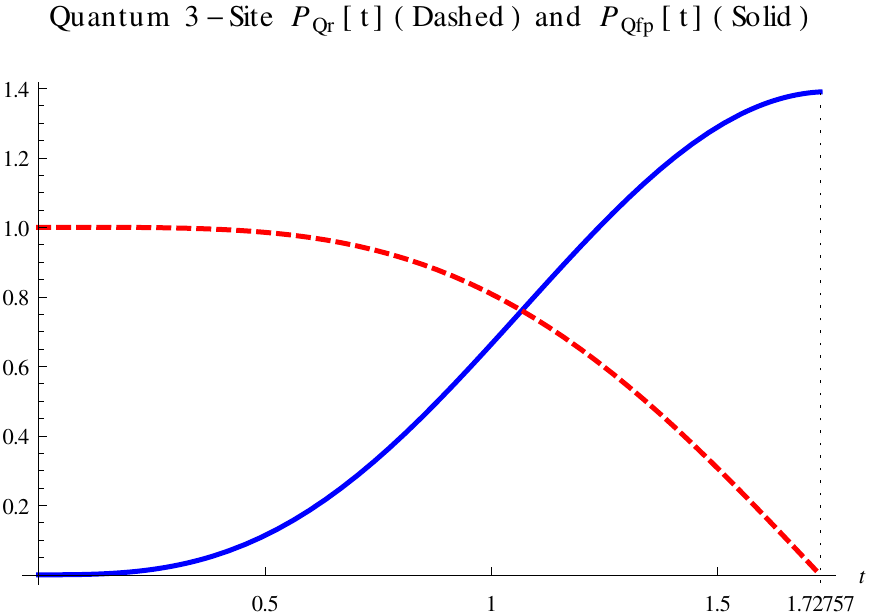}
\caption{\small{Quantum 3-site system: Plots of the restricted probability
$P_r^Q \forr{\omega}{t}{1}{0}$ (dashed line) and the first-passage probability density 
$P_{fp}^Q \forr{\omega}{t}{1}{0}$ (solid line) against $ t$, with $T = 1.72757$.}}\label{f3}
\end{figure}

The above treatment can be readily extended to the other few-site cases with appropriate choices for the subspaces $\omega$ and $\bar{\omega}$. Thus, for the 3-site quantum system $\{ \kett{1}, \kett{2}, \kett{3} \}$ with the tight-binding Hamiltonian
\be \label{e9}
\mathcal{H}_3~=~-\oter{1}{2}-\oter{2}{1}-\oter{2}{3}-\oter{3}{2}~~,
\ee
let us consider the case in which $\omega \equiv \{\kett{1}, \kett{2} \}$
and $\bar{\omega} \equiv \kett{3}$. When the system is initiated  at time $t=0$ at site $\kett{1}$ of $\omega $, the following relevant \textit{unrestricted} probabilities are obtained: 
\be \label{e10}
\begin{split}
P_u^Q \forr{\omega}{t}{1}{0}
&=\modsq{ \braa{1} e^{- i \mathcal{H}_3 t} \kett{1} } + \modsq{ \braa{2} e^{- i \mathcal{H}_3 t} \kett{1}} \\
&=\frac{3-\cos \left(\sqrt{2}~t\right)}{2} \cos ^2\left(\frac{t}{\sqrt{2}}\right)~~,
\end{split}
\ee
\be \label{e11}
\begin{split}
P_{u}^Q \forr{\omega}{t}{3}{t^\prime}
&=\modsq{ \braa{1} e^{- i \mathcal{H}_3 (t-t^{\prime}) } \kett{3} } + \modsq{ \braa{2} e^{- i \mathcal{H}_3 (t-t^{\prime})} \kett{3} } \\
&=\frac{3+\cos \left(\sqrt{2}~(t-t^\prime)\right)}{2} \sin ^2\left(\frac{t-t^\prime}{\sqrt{2}}\right).
\end{split}
\ee
In terms of these, we have the exact solutions given below (with the numerical coefficients rounded off), as plotted in Fig \ref{f3}.
\be \label{e12}
P_r^Q \forr{\omega}{t}{1}{0}=1.132 \cos(0.915~t) - 0.132 \cos(2.676~t),
\ee
and
\be \label{e13}
P_{fp}^Q \forr{\omega}{t}{1}{0}=1.036 \sin(0.915~t) - 0.354 \sin(2.676~t).
\ee
\begin{figure}
\centering
\includegraphics[width=.45\textwidth, clip=true,trim=0pt 0pt 0pt 14pt]{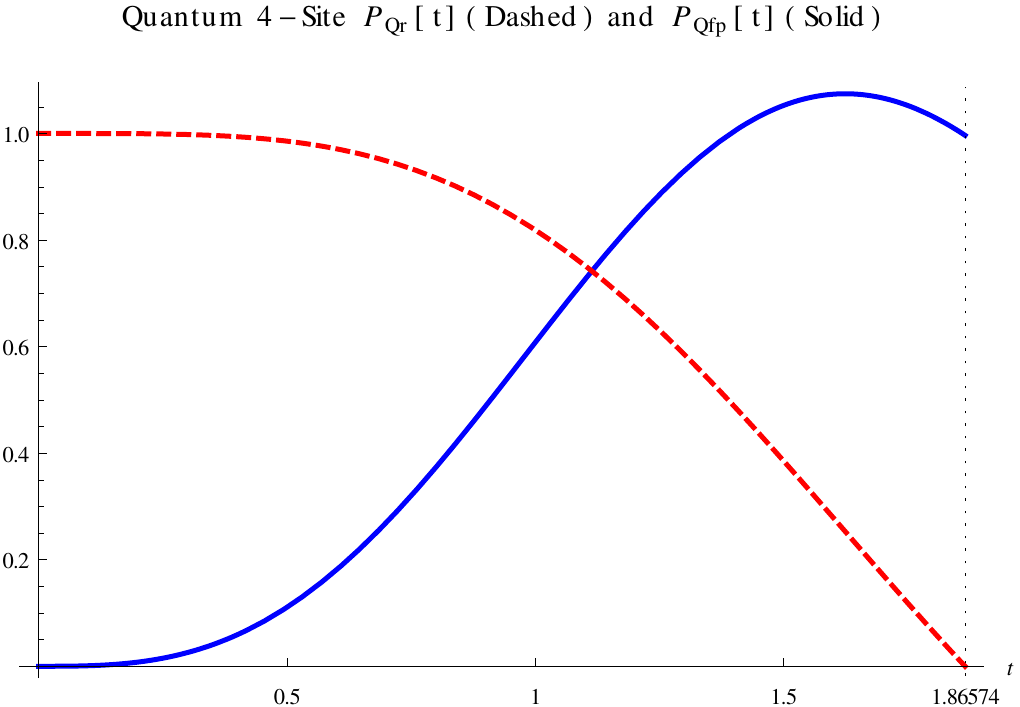}
 \caption{\small{Quantum 4-site system: Plots of the restricted probability
$P_r^Q    \forr{\omega}{t}{1}{0}$ (dashed line) and the first-passage probability density
$P_{fp}^Q \forr{\omega}{t}{1}{0}$ (solid line) against $ t$, with $T = 1.86574.$}} \label{f4}
\end{figure}

Similarly, the results obtained for a 4-site system $\{\kett{1},\kett{2},\kett{3},\kett{4}\}$, with $\omega \equiv \{\kett{1}, \kett{2} \}$ and \mbox{$\bar{\omega} \equiv \{ \kett{3}, \kett{4} \}$}, initiated at $\nu \equiv \kett{1}$, are given below and plotted in Fig. \ref{f4}.
\be \label{e14}
\begin{split}
P^Q_{r}\forr{\omega}{t}{1}{0}~=
&~0.954~\cos (0.754~t) + 0.131 \cos (1.261~t) \\
&-0.085 \cos (2.973~t),
\end{split}
\ee
and,
\be \label{e15} 
\begin{split}
P^Q_{fp}\forr{\omega}{t}{1}{0}~=
&~0.719 \sin (0.754~t) + 0.165 \sin (1.261~t) \\
& - 0.253 \sin (2.973~t)
\end{split}
\ee
It should be noted here that for all the few-site cases treated above, the three physical conditions (A, B and C) listed previously are indeed satisfied, and the time T defining the domain for the Q-fpt problem is found to remain finite. Also, by comparing the results obtained for the 4-site and the 3-site problems, it is observed that the addition of extra distant sites in the complementary subspace $\bar\omega$ does indeed modify the Q-fpt distribution. This is quite contrary to the case of diffusive C-fpt where it does not. This is essentially a manifestation of the so-called \textit{non-locality} of quantum mechanics, orchestrated by the \textit{virtual excursions} across the boundary $(\sigma)$.

Finally, we consider the case of an infinite 1D lattice with the tight-binding Hamiltonian
\be \label{e16}
\mathcal{H}_\infty~=~-\sum_{n=-\infty}^\infty \left(\kett{n+1}\braa{n}~+~\kett{n}\braa{n+1}\right),
 \ee
where we choose $\omega \equiv \{0,-1,-2,-3\ldots,-\infty\}$ and $\bar{\omega} \equiv \{1,2,3,\ldots,+\infty\}$. The system is initiated at site $\kett{0}$ at time $t=0$. In this case, the \sch integral equation 
\be \label{e17}
\begin{split}
P_u^Q\forr{\omega}{t}{0}{0}
&=P_r^Q\forr{\omega}{t}{0}{0} \\
&-\int_0^t \dd t^\prime~P_u^Q\forr{\omega}{t}{1}{t^\prime}
~\frac{\partial}{\partial t^\prime}
P_r^Q\forr{\omega}{t}{0}{0},
\end{split}
\ee
reduces to give simpler expressions for the Laplace-transforms $(\mathcal{L})$ of $P^Q_{r}$ and $P^Q_{fp}$ :
\be \label{e18}
\mathcal{L}\sqbr{P_r^Q\forr{\omega}{t}{0}{0}} =
\frac{\mathcal{L}\sqbr{P_u^Q\forr{\omega}{t}{0}{0} - P_u^Q\forr{\omega}{t}{1}{0}}}
{1-s\mathcal{L}\sqbr{P_u^Q\forr{\omega}{t}{1}{0}}},
\ee 
\be \label{e19}
\mathcal{L}\sqbr{P_{fp}^Q\forr{\omega}{t}{0}{0}}
=\frac{1-s\mathcal{L}\sqbr{{P_u^Q\forr{\omega}{t}{0}{0}}}}
{1-s\mathcal{L}\sqbr{P_u^Q\forr{\omega}{t}{1}{0}}}.
\ee
Now, with the help of the time-Laplace and the lattice-Fourier transforms, we solve for the \textit{unrestricted} unitary evolutions of the wavefunction $\kett{\psi(t)} = \sum_{n=-\infty}^{\infty} a_n(t) \kett{n}~$, given by the \textit{time-dependent \sch equation}
\be \label{e20}
-i\dot{a_n}(t)~=~a_{n-1}(t)~+~a_{n+1}(t)~~,
\ee
for the two initial conditions: (I) for $~a_n(0) = \delta_{n,0}$ giving
\be \label{e21}
P_u^Q\forr{\omega}{t}{0}{0} \equiv \sum_{n=0}^{-\infty}\modsq{a^{\rm I}_n(t)} = \frac{(1+J^2_{0} (2t))}{2} ~,
\ee
and, (II) for $~a_n(0) = \delta_{n,1}$ giving
\be \label{e22}
P_u^Q\forr{\omega}{t}{1}{0} \equiv \sum_{n=0}^{-\infty}\modsq{a^{\rm II}_n(t)} = \frac{(1-J^2_{0} (2t))}{2}~,
\ee
where $J_0$ denotes the Bessel function of order zero. 
Here the superscripts (I and II) on the amplitudes $a_n(t)$ denote the corresponding initial conditions. Upon substituting from eqs. (19) and (20) back into eqs(16) and (17), we obtain
\be \label{e23}
\mathcal{L}\sqbr{P_r^Q\forr{\omega}{t}{0}{0}}
~=~\frac{4}{s} ~\frac{{\rm K}\left(-\frac{16}{s^2}\right)}{\pi +2 {\rm K}\left(-\frac{16}{s^2}\right)}~~,
\ee
\be \label{e24}
\mathcal{L}\sqbr{P_{fp}^Q\forr{\omega}{t}{0}{0}}
~=~\frac{\pi-2~{\rm K}\left(-\frac{16}{s^2}\right)}{\pi +2~{\rm K}\left(-\frac{16}{s^2}\right)}~~,
\ee
where, $K(x)$ is the complete elliptic integral of the first kind, and, we have used the identity $\mathcal{L}\sqbr{J_0^2(at)} = \frac{2}{\pi s}~\text{K}\parnth{-\frac{4a^2}{s^2}}$ \cite{30,31}.

\begin{figure}
\centering
\includegraphics[width=.45\textwidth, clip=true,trim=0pt 0pt 0pt 24pt]{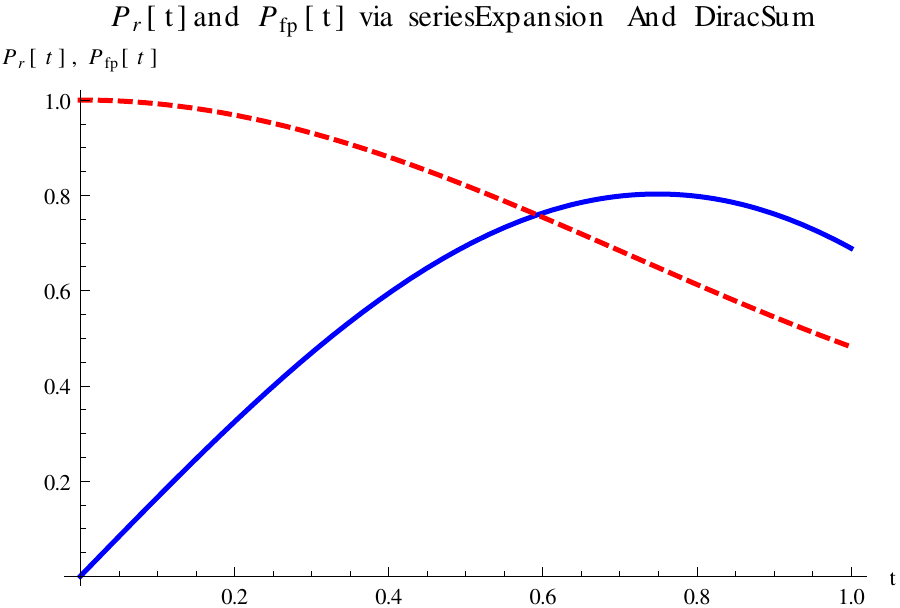}
 \caption{\small{Quantum infinite lattice system: Small-time behaviors of the restricted probability
$P_r^Q    \forr{\omega}{t}{0}{0}$ (dashed line) and the first-passage probability density
$P_{fp}^Q \forr{\omega}{t}{0}{0}$ (solid line) against $t$.}} \label{f5}
\end{figure}

Equations \eqr{e23} and \eqr{e24} cannot be written in terms of elementary functions. From the large $s$ behavior of the Laplace transforms in eqs. \eqr{e23} and \eqr{e24}, however, we have obtained the small-time behavior of the required \textit{restricted} and \textit{first-passage} probabilities $P^Q_r$ and $P^Q_{fp}$ (using a professional software package like \cite{31}). These are plotted in Fig. \ref{f5}.
As can readily be seen from the plots, in this limited small time regime, $P_r^Q$ remains positive while falling monotonically, whereas $P^Q_{fp}$ remains positive, though not monotonic. We emphasize that both of these trends are physically valid.

In conclusion, we have presented an extension of the \sch integral equation for the classical first-passage time (C-fpt) probability density to the case of quantum first-passage time(Q-fpt) probability density. We would like to emphasize that this quantum extension is only for the class of the N-site tight-binding Hamiltonians considered here by us. These models are characterized by the presence of a unique pair of sites astride the boundary (post/marker). This pair of sites has been aptly referred to as the \textit{door-way} in the present work. For these models, all quantum interference effects of alternatives are contained in the two unrestricted quantum probabilities $P_u^Q (t)$ taken in conjunction with the restricted quantum probability  $P_r^Q (t)$, which gets determined implicitly by the integral equation (e.g., \eqr{e4}). This is so because the event of a first passage through the door-way is a unique space-time point that all quantum amplitudes from $\omega$  to $\bar{\omega}$ must pass through $-$ 
this, however, subtends no phase-shifts of relevance at that space-time point. This, of course, is not the case when more than one door-ways are present on the border between $\omega$ and $\bar{\omega}$. 

With this extension, we have calculated Q-fpt probability densities for some cases of N-site model systems, i.e., 2-site (qubit), 3-site and the 4-site. We have also treated the case of an infinite 1D tight-binding Hamiltonian system. The thusly calculated Q-fpt probability densities satisfy the necessary physical conditions of positivity and normalization to unity over the time-domain of definition of the problem. The latter turns out to be finite unlike for the case of C-fpt. Additionally, they carry another distinct quantum signature: even the distant sites beyond the boundary affect the Q-fpt distributions. This is quite unlike the case for C-fpt. This point is clearly brought out when we compare the results for the various cases of the N-site tight-binding Hamiltonian systems. This is understandable as a consequence of the virtual quantum excursions across the boundary.

We hope that the present approach opens up a novel view upon this important field of Q-fpt problems whose solutions have so far eluded us. More specifically, this approach should be of special relevance to the quantum phenomena of inter-biomolecular/nanostructural electron transfers \cite{7, 14, 17,18}, where such \textit{door-way} states are expected to exist.

\begin{acknowledgments}
One of us (RV) thanks Prof Anil Kumar and acknowledges DST-0955 project (CQIQC) at IISc for support during the course this work. RV also thanks RRI for its hospitality.
\end{acknowledgments}

\end{document}